# Direct observation of symmetry-breaking in a 'ferroelectric' polar metal

Emma Berger[1,2,*], Sasawat Jamnuch[3], Can Uzundal[1,2], Clarisse Woodahl[4], Hari Padmanabhan[5], Angelique Amado[1,2], Paul Manset[6], Yasuyuki Hirata[7], Iwao Matsuda[8,9], Venkatraman Goplan[5], Yuya Kubota[10,11], Shigeki Owada[10,11], Kensuke Tono[10,11], Makina Yabashi[10,11], Youguo Shi[12], Craig Schwartz[13], Walter Drisdell[13,14], John Freeland[15], Tod Pascal[3,16,17], Michael Zuerch[1,2,18,*]

1. Department of Chemistry, University of California, Berkeley, California 94720, USA
2. Materials Sciences Division, Lawrence Berkeley National Laboratory, Berkeley, California 94720, USA
3. ATLAS Materials Science Laboratory, Department of Nano Engineering and Chemical Engineering, University of California, San Diego, La Jolla, California, 92023, USA
4. Department of Materials Science and Engineering, The Pennsylvania State University, University Park, Pennsylvania 16801, USA
5. University of Florida, Gainesville, FL 32611, USA
6. Ecole Normale Superieure de Paris, Paris, France
7. National Defense Academy of Japan, Yokosuka, Kanagawa 239-8686, Japan
8. Institute for Solid State Physics, The University of Tokyo, Kashiwa, Chiba 277-8581, Japan
9. Trans-scale Quantum Science Institute, The University of Tokyo, Bunkyo-ku, Tokyo 113-0033, Japan
10. RIKEN SPring-8 Center, Sayo 679-5148, Japan
11. Japan Synchrotron Radiation Research Institute (JASRI), Sayo 679-5198, Japan
12. Institute of Physics, Chinese Academy of Sciences, Beijing 100190, China
13. Chemical Sciences Division, Lawrence Berkeley National Laboratory, Berkeley, California 94720, USA
14. Joint Center for Artificial Photosynthesis, Lawrence Berkeley National Laboratory, Berkeley, California 94720, USA
15. Advanced Photon Source, Argonne National Laboratory, Argonne, IL 60439, USA
16. Materials Science and Engineering, University of California San Diego, La Jolla, California, 92023, USA
17. Sustainable Power and Energy Center, University of California San Diego, La Jolla, California, 92023, USA
18. Fritz Haber Institute of the Max Planck Society, Berlin, Germany

**Ferroelectric materials contain a switchable spontaneous polarization that persists even in the absence of an external electric field. The coexistence of ferroelectricity and metallicity in a material appears to be illusive, since polarization is ill-defined in metals, where the itinerant electrons are expected to screen the long-range dipole interactions necessary for dipole ordering. The recent discovery of the polar metal, lithium osmate ($LiOsO_3$) was therefore surprising[1] and has generated a renewed interest in searching for new polar metals motivated by the prospects of nanocapacitors, unconventional superconductors, topological spin currents, anisotropic upper critical fields, and nano-heterostructuredmultiferroics[2–5]. Previous studies have suggested that the coordination preferences of the lithium atom play a key role in stabilizing the polar metal phase of $LiOsO_3$, but a thorough understanding of how polar order and metallicity can coexist remains elusive[2,6,7]. Here, we use extreme ultraviolet second harmonic generation (XUV-SHG) as novel technique to directly probe the broken inversion-symmetry around the Li atom. Our results agree with previous theories that the primary structural distortion that gives rise to the polar metal phase in $LiOsO_3$ is a consequence of a sub-Angstrom Li atom displacement along the polar axis. A remarkable agreement between our experimental results and *ab initio* calculations provide physical insights for connecting the nonlinear response to unit-cell spatial asymmetries. It is shown that XUV-SHG can selectively probe inversion-breaking symmetry in a bulk material with elemental specificity. Compared to optical SHG methods, XUV-SHG fills a key gap for studying structural asymmetries when the structural distortion is energetically separated from the Fermi surface. Further, these results pave the way for time-resolved probing of symmetry-breaking structural phase transitions on femtosecond timescales with element specificity.**

Broken symmetries are of cardinal importance for emergent phenomena across nearly all scientific disciplines with magnetism, superconductivity, ferroelectricity, and the formation of DNA being just a few representative examples. A complete and fundamental understanding of the structure-function relationships that arise as a result of broken symmetries remains elusive but will be key for novel breakthroughs in these areas and beyond. A particularly interesting material class with a non-intuitive symmetry breaking, namely, the coexistence of polarity and metallicity, are 'ferroelectric-like' polar metals. Though it was first predicted over 50 years ago that polar metals could form through a 2nd order phase transition, the first experimentally realized 'ferroelectric-like' metal, lithium osmate ($LiOsO_3$) was discovered only recently[1,8]. The search for polar metals has since expanded, motivated by prospects of Mott multiferroics[3], electrodes in ferroelectric nanocapacitors[6], nonlinear optical media[9], polar superconductors[4,5,10], and thermoelectric devices[11]. Recent works predicting asymmetric hysteresis in $LiOsO_3$ thin films and revealing ferroelectric switching in a two-dimensional metal[12] have generated significant recent excitement for the practical possibility of polar metal device integration in low-dimensional platforms. Despite the numerous attempts to explain the co-existence of polarity and metallicity, there are still many unanswered questions as to how polar order can be stabilized in a metal.

Since its discovery, $LiOsO_3$ has become the prototypical polar metal. In this perovskite-structured material ($ABO_3$), a continuous order-disorder phase transition occurs at a critical temperature $T_c$ = 140 K, where it transitions from an $R\bar{3}c$ nonpolar metallic to an R3c polar metallic phase through the loss of inversion symmetry (Fig. 1a). Neutron diffraction[1], Raman[13], and optical spectroscopy[14] have shown that the $A_{2u}$ soft phonon mode responsible for the transition involves a coordinated 0.5 Å displacement of Li atoms along the polar c-axis[13]. To explain these observations, several theories have been developed[7,15–18]. The decoupled electron mechanism hypothesis proposed by Puggioni *et al.* postulates that the soft phonon is responsible for driving the 'ferroelectric' transition are energetically decoupled from the electrons at the Fermi level responsible for the metallicity. In this picture, ineffective electron screening of long-range dipolar couplings stabilize the polar phase[11,14,19]. Benedek *et al.* propose a related mechanism that emphasizes short-range, atomic detail[2]. Here, ion-size mismatch effects and the local coordination preferences of the A-site Li atom are thought to drive the phase transition. With short-range interactions at the forefront, the itinerant electrons that are expected to screen the dipole-dipole forces are thus of secondary importance. In the high-temperature phase, the Li atom is coordinated to nine nearby oxygen atoms via three short Li-O bonds and six long Li-O bonds. On the other hand, signatures of the polar metal phase include $OsO_6$ octahedral rotations that shorten three of the longer Li-O bonds, and Li atoms displacements that shorten the three short Li-O bonds. The overall effect of these structural changes is to octahedrally coordinate the Li atom in the polar metal phase[2] (Fig. 1b). The coordinated motion of all Li atoms in the same direction is also thought to minimize the number of faces Li octahedra share with Os octahedra and that anisotropic Coulomb screening around the Li atoms to interact through unscreened channels[7,15]. Hence, the Li-coordination environment plays a determining role in the stabilization of the polar metal phase, but a direct experimental probe of the inversion-broken symmetry around the Li environment is lacking. The most common probe of broken inversion symmetry is optical second harmonic generation (SHG) spectroscopy, which has been applied previously to examine the nature of Os-O bonding in $LiOsO_3$ [20]. However, not only is optical SHG non-element specific, but a partial density of states (DOS) analysis reveals overwhelming contributions Os 5d and O 2p character around the Fermi level (Fig. 1c & SI Section 6). Indeed, this observation forms the basis for the aforementioned DEM hypothesis, but the implication here is that optical SHG is not sensitive to the Li-coordination environment.

To gain insight into the nature of the Li-coordination environment in the polar phase, we turn to extreme ultraviolet second harmonic generation (XUV-SHG) spectroscopy, a newly developed method viable for studying bulk-phase non-centrosymmetric materials[21], surfaces[22,23], and buried interfaces[24] in either the XUV or the soft x-ray (SXR) regimes. Hard x-ray (HXR) SHG has also been attempted before[25], but the HXR wavelengths are on the order of unit-cell dimensions, thus making even centrosymmetric materials appear non-centrosymmetric. In SXR- or XUV-SHG spectroscopy, the incident x-ray beam is resonant with a core-to-valence transition, such that the resulting signal is highly sensitive to the element-specific electronic structure. The application of XUV-SHG spectroscopy to such short wavelengths, though, requires the use of XUV photon sources capable of achieving high pulse intensities and coherence to drive the nonlinear matter response, for which x-ray free electron lasers (XFEL) are uniquely well-suited. The inherent intensity fluctuations in XFEL beams allow for the measurement of the intensity-dependent XUV-SHG signal. The second-order response can then be measured via

$$I(2\omega) \propto |\chi_{\text{eff}}^{(2)}(2\omega; \omega + \omega)|^2 \, I(\omega)^2 \tag{1}$$

where $\chi_{\text{eff}}^{(2)}$ is the effective nonlinear susceptibility, dependent on crystal cut and experimental geometry, the incident beam is at a frequency-dependent intensity, $I(\omega)$, and SHG emission occurs with a frequency of $2\omega$. Here, the dielectric environment around Li in LiOsO$_3$ below $T_c$ is directly probed by tuning the incident XFEL photon energy to be half-resonant with energies around the Li K-edge at -56.6 eV (Fig. 1c). The energy-dependent $\chi_{\text{eff}}^{(2)}$ is extracted according to equation (1) and *ab initio* density functional perturbation theory (DFPT) simulations are used to relate the nonlinear response to the broken inversion symmetry. Our results directly show that the broken inversion symmetry in LiOsO$_3$ are a result of asymmetric distortions involving Li atoms, rather than those involving Os-O displacements.

**Experimental Results**

XUV-SHG spectroscopy was performed on samples of LiOsO$_3$ in its polar metal phase at $T$=62 K at the SPring-8 Angstrom Compact Free Electron Laser (SACLA) in Japan at the BL1 beamline[26]. A 30 fs p-polarized XUV beam was incident at 45° with respect to the surface normal onto the [120] plane of the LiOsO$_3$ sample. By tuning the incident photon energy to energies ranging from 28-33 eV, the incident photon was half-resonant with the Li 1s (-56.6 eV) level and close-by Os 4f (-61.6 eV) and 5p (-60.2 eV) levels, thus requiring two incident photons to access the valence states (Fig. 1d). XUV-SHG spectra were collected by energy-dispersing the outgoing beam with a grating to simultaneously observe the fundamental and second harmonic beams with a fluorescent screen-coated micro-channel plate detector (Fig. 1e). The $\chi_{\text{eff}}^{(2)}$ spectrum was then extracted using equation (1) after binning and averaging spectra based on the photon energy and intensity of the driving pulse. The inherent jitter of the XFEL, both in frequency and intensity, allowed for a high-fidelity extraction of $\chi_{\text{eff}}^{(2)}(2\omega)$ with significantly more data points as compared to Refs. 21–24. At each photon energy, a quadratic function was fit to a plot of $I(\omega)$ vs. $I(2\omega)$, where the second-order fit coefficient is modulo $|\chi_{\text{eff}}^{(2)}(2\omega)|^2$, with $\chi_{\text{eff}}^{(2)}(2\omega)$ a rank-three tensor of four independent components ($\chi_{zzz}^{(2)}, \chi_{xxz}^{(2)}, \chi_{zxx}^{(2)}$ and $\chi_{xxy}^{(2)}$) due to space group symmetry. In order to relate the measured nonlinear susceptibility to unit cell structure, a full tensor analysis considering the experiment geometry and crystal cut was performed to arrive at a mathematical expression for $\chi^{(2)}_{\text{eff}}$. Further details of the experimental setup, data analysis, and sample preparation are described in the *Methods* section and *Sections S1-4* in the *Supplementary Materials*.

According to selection rules for SHG, electrons residing in Li 1s core states can transition to valence states of Os 5d character, whereas the Os 5p and Os 4f semi-core states can transition to valence states of O 2p character (Fig. 1c). The approximately 2-eV-wide gap in the DOS above the Fermi level ($E_F$) has been previously assigned to the crystal-field splitting in Os 5d-like orbitals, but the nonzero DOS directly at $E_F$ is consistent with the metallic character of LiOsO$_3$[17].

To gain insight how the measured spectra relate to the Li coordination environment, theoretical calculations of the LiOsO$_3$ linear absorption were performed using the *exciting* full potential all electron augmented linearized planewave software package based on first principles density function theory (DFT)[27]. Two LiOsO$_3$ periodic cell structures, corresponding to the nonpolar and polar phases, were used. The Brillouin zone was sampled with a 15x15x15 *Γ*-point centred *k*-point grid with the local density approximation functional[28]. DFPT simulations within the random phase approximation were used to access excited states of the system. The formalism outlined by Sharma *et al.* as implemented within *exciting* was used calculate the four active $\chi^{(2)}_{\text{ijk}}$ tensor elements (Fig 2a-d), with the aforementioned *k*-point grid set again[29]. 120 empty states were included in the ground state calculation to account for the excited state at double the energy of the Li 1s core state. The background signal, which is the response from the valence electrons and is proportional to ~1/energy, was subtracted to obtain the effective susceptibility from the core state. The $\chi^{(2)}_{\text{eff}}$ was next calculated as a function of varying Li position along the polar axis, *α*, and compared to the experimental results (Fig. 2e). Simulations with OsO$_6$ displacements were also performed (see *Supplementary Information Section S7)* showing only little effect on the resulting XUV-SHG spectra. In addition, the nonlinear susceptibility was calculated with the exclusion of Li 1s states (see *Supplementary Information Section S8),* the results of which confirm that the nonlinear response within the energy range probed is predominantly due to transition from Li 1s core states to states above the Fermi level. Finally, the DFT calculations were repeated using the Vienna *ab initio* software package (VASP[30]) using a projector augmented wave (PAW) approach and a plane wave basis set of up to 400 eV to visualize the Kohn-Sham equation-generated charge densities within the LiOsO$_3$ unit cell. Shown in Fig. 3 are projections of this charge density surface onto the [110] plane through the middle of the hexagonal unit cell.

**Discussion**

The experimentally determined $\chi_{\text{eff}}^{(2)}$ shown in Fig. 2e has several features to point out. DFPT calculations indicate that the XUV-SHG signal in the 56-59 eV range is overwhelmingly a result of transitions from Li 1s states. This is supported by the observation in Fig. 2e that the spectral feature centred at 58 eV is highly sensitive to Li atom position within the unit cell. The remarkable agreement between theory and experiment confirms that XUV-SHG is indeed sensitive to such symmetry breaking. Also of note is the lack of SHG response in the energy range above 60 eV that could result from electronic transitions from Os semi-core to O 2p states, which likely results from that fact that inversion symmetry is largely maintained around Os atoms, as will be discussed further below.

Next, we turn to an analysis of how the XUV-SHG spectrum varies as a function of Li displacement $α$. The main feature at 58 eV increases in magnitude with asymmetry in the Li-O dipoles along the polar axis, resulting from one of two factors: a) the mere increase in the asymmetric distortion enhances the XUV-SHG signal, or, b) as Li-ions displace towards OsO$_6$ octahedra, the aforementioned Li-O "long bonds" get shorter, leading to increased hybridization between orbitals of Li 2s and Os 2p character. This, in turn, gives more s-character to O 2p orbitals, thus opening up more selection-rule-allowed states for Li 1s electrons to transition into. The latter possibility is supported by DFT simulations since increased hybridization should lead to a downward shift in energy, which is observed in the appearance of a spectral peak at 57.5 eV as $α$ increases.

To gain further insight into the inversion symmetry-breaking process, projections of the calculated charge density surface onto the [110] plane of the real space unit cell were examined. Shown in Fig. 3 are slices of the charge density surfaces cut along the [110] plane in both the polar and nonpolar phases of LiOsO$_3$. It is immediately apparent that the polar phase is characterized by broken inversion symmetry around the Li ions as evidenced by the unequal relative displacements of each of the Li atoms to the central Os atom in the polar phase.

The plots in Fig. 3 provide further qualitative physical intuition behind spectral features of the tensor elements shown in Figs. 2a-d. Here, it should be noted that the $i^{th}$ index in $\chi^{(2)}_{ijk}$ corresponds to the Cartesian direction along which dipole oscillations are generated at with a frequency of $2\omega$ by driving electric fields along the $j^{th}$ and $k^{th}$ directions. The relatively sharp feature at 58 eV observed in the calculated spectra for $\chi^{(2)}_{zzz}$ and $\chi^{(2)}_{zxx}$ can be attributed to the polar displacement of Li atoms along the z-axis as the primary contributor to the broken inversion symmetry in LiOsO$_3$. It can also be seen that $\chi^{(2)}_{zzz}$ and $\chi^{(2)}_{xxz}$ increase in magnitude, but with a relatively unchanged spectral shape as $\alpha$ increases. In contrast, $\chi^{(2)}_{zxx}$ exhibits a nontrivial $\alpha$-dependence. These observations can be contextualized by comparing the nonpolar and polar charge densities in Fig. 3. Here, it is apparent that driving with a fundamental pulse along the z-direction will increase the asymmetry along the polar axis. Increasing $\alpha$ merely increases the amplitude of these oscillations. On the other hand, an incident laser beam polarized along the x-direction will alter the electron density with respect to the already present Li displacement in a complex way. Lastly, it is of note that $\chi^{(2)}_{xxy}$ changes minimally with $\alpha$. This can again by explained by the observation that there are minimal polar displacements within the x-y plane for all atomic environments within the unit cell.

In summary, it is shown that XUV-SHG can selectively probe inversion-breaking symmetry in a bulk material with elemental specificity. Compared to optical SHG methods, XUV-SHG fills a key gap for studying structural asymmetries when the structural distortion is energetically separated from the Fermi surface. With a remarkable agreement between theory and experiment in LiOsO$_3$ it is found that the presence of XUV-SHG at the Li K-edge and lack thereof at Os semi-core edges support previous claims that the stability of the polar phase in LiOsO$_3$ stems from the Li displacement along the polar axis. Furthermore, a comparison of calculated charge density plots and theoretical simulations of the how the nonlinear susceptibility tensor elements vary as a function of Li displacement provide a qualitative picture for how microscopic asymmetries on unit cell length scales determine the nonlinear response. Our findings pave the way to study structural phase transitions and critical nonequilibrium phenomena with elemental specificity and femtosecond time resolution in other geometric ferroelectrics, multiferroics and complex, multi-component materials.

**Methods**

*Experimental setup*

The FEL beam was passed through a 0.8 µm Al filter used to attenuate the beam to prevent sample damage before hitting the [120] plane of $LiOsO_3$ at an angle of 45° with respect to the surface normal and with a focused spot size with full-width-half-maximum of 50 µm. The resultant XUV-SHG signal and weak reflected fundamental FEL beams are spectroscopically analysed. Both beams are incident at 87° with respect to the surface normal onto a 1200 groove/mm grating (30-002, Shimadzu) following a 200 µm slit. A micro-channel plate (MCP) detector (F2224, Hamamatsu Photonics) coated with CsI to image the SXR spectrum (IPX-VGA120-LMCN, Imperx Inc.) is used for detection. Experiments were carried out below $T_c$ at 62 K with ten different incident energies ranging from 28-33 eV. The intensity of each shot was calibrated by measuring the photoionization of Ar in a gas cell prior to hitting the sample where it was previously shown that 90% of the beam incident on the gas is transmitted onto the sample[26]. No signal above signal to noise was detected above the transition temperature $T_c$=140 K.

*Data Processing*

Each shot was registered on the MCP detector as a 2D image containing the fundamental peak and the second harmonic peak at frequencies of $\omega$ and $2\omega$, respectively. Shots of poor quality were identified as those with a full-width half-maximum (FWHM) at the fundamental frequency greater than 1.5 standard deviations away from the average FWHM and discarded. The remaining spectra were background corrected, binned, and averaged according to the intensity of the incident laser pulse. Within each bin, the average intensity of the fundamental and second harmonic peaks were extracted. A quadratic function was fit to the $I(\omega)$ vs. $I(2\omega)$ plot at each incident energy to extract the nonlinear susceptibility according to equation (1) as shown in Figure S1.1. Further details on the spectral analysis can be found in Section S1 of the supplementary materials. This process was then repeated at all incident energies to produce a modulo $\chi^{(2)}$ spectrum. To compare experiment with theory, a 35.4 eV rigid shift in the TDDFT energy axis was determined to be necessary. The magnitude of the shift was chosen by finding the best fit to the experimental results. Such a treatment has been shown to be a valid approximation under the assumption that $\chi^{(2)}$ spectrum lineshapes are the same at the second-order response of the fundamental and first-order response of the SHG and differ only by an additive offset.[31]

*Sample*

Single crystals of $LiOsO_3$ were prepared using a solid-state reaction at high pressure, as outlined in a previous publication[1]. The crystals were hand-polished using calcined alumina polishing paper of roughness 0.3 µm. Electron back-scattering diffraction was then used to orient the crystals and confirm crystallinity.

**Acknowledgements:** M. Z., C. S. and A. A. acknowledge support by the Max Planck Society (Max Planck Research Group). M. Z. acknowledges support by the Federal Ministry of Education and Research (BMBF) under "Make our Planet Great Again – German Research Initiative" (Grant No. 57427209 "QUESTforENERGY") implemented by DAAD. H.P., J.W. F, and V. G. acknowledge Department of Energy grant number DE SC-0012375. W.D. acknowledges support from the Joint Center for Artificial Photosynthesis, a DOE Energy Innovation Hub, supported through the Office of Science of the U.S. Department of Energy, under Award No. DE-SC0004993. Measurements were performed at BL1 of SACLA with the approval of the Japan Synchrotron Radiation Research Institute (JASRI) (Proposal No. 2019B8066). This work was supported by the SACLA Basic Development Program 2018-2020. The authors would like to acknowledge the supporting members of the SACLA facility. Additional measurements were performed at beamline 6.3.2 of the Advanced Light Source, a U.S. DOE Office of Science User Facility under contract no. DE-AC02-05CH11231. This research used resources of the National Energy Research Scientific Computing Center, a DOE Office of Science User Facility supported by the Office of Science of the U.S. Department of Energy under Contract No. DE-AC02-05CH11231. This work also used the Extreme Science and Engineering Discovery Environment (XSEDE), which is supported by National Science Foundation grant number ACI-1548562. We are grateful for input and discussion with David Attwood and Ramamoorthy Ramesh.

**Competing Interests:** The authors declare that they have no competing financial interests.

**Correspondence:** Correspondence and requests for materials should be addressed to E. Berger (emma_berger@berkeley.edu) and M. Zuerch (mwz@berkeley.edu).

**References**

1. Shi, Y. *et al.* A ferroelectric-like structural transition in a metal. *Nat. Mater.* **12**, 1024–1027 (2013).

2. Benedek, N. A. & Birol, T. 'Ferroelectric' metals reexamined: Fundamental mechanisms and design considerations for new materials. *J. Mater. Chem. C* **4**, 4000–4015 (2016).

3. Puggioni, D., Giovannetti, G., Capone, M. & Rondinelli, J. M. Design of a Mott Multiferroic from a Nonmagnetic Polar Metal. *Phys. Rev. Lett.* **115**, 1–6 (2015).

4. Edelstein, V. M. Magnetoelectric effect in polar superconductors. *Phys. Rev. Lett.* **75**, 2004–2007 (1995).

5. Enderlein, C. *et al.* Superconductivity mediated by polar modes in ferroelectric metals. *Nat. Commun.* **11**, 4852 (2020).

6. Puggioni, D., Giovannetti, G. & Rondinelli, J. M. Polar metals as electrodes to suppress the critical-thickness limit in ferroelectric nanocapacitors. *J. Appl. Phys.* **124**, (2018).

7. Liu, H. M. *et al.* Metallic ferroelectricity induced by anisotropic unscreened Coulomb interaction in LiOsO3. *Phys. Rev. B - Condens. Matter Mater. Phys.* **91**, 2–7 (2015).

8. Anderson, P. W. & Blount, E. I. Symmetry considerations on martensitic transformations: 'ferroelectric' metals?. *Phys. Rev. Lett.* **14**, 532 (1965).

9. Edelstein, V. M. Features of light reflection off metals with destroyed mirror symmetry. *Phys. Rev.*

*B - Condens. Matter Mater. Phys.* **83**, 1–4 (2011).

10. Edelstein, V. M. Magnetoelectric effect in dirty superconductors with broken mirror symmetry. *Phys. Rev. B - Condens. Matter Mater. Phys.* **72**, 1–4 (2005).

11. Puggioni, D. & Rondinelli, J. M. Designing a robustly metallic noncenstrosymmetric ruthenate oxide with large thermopower anisotropy. *Nat. Commun.* **5**, 1–9 (2014).

12. Fei, Z. *et al.* Ferroelectric switching of a two-dimensional metal. *Nature* **560**, 336–339 (2018).

13. Jin, F. *et al.* Raman interrogation of the ferroelectric phase transition in polar metal LiOsO3. *Proc. Natl. Acad. Sci. U. S. A.* **116**, 20322–20327 (2019).

14. Laurita, N. J. *et al.* Evidence for the weakly coupled electron mechanism in an Anderson-Blount polar metal. *Nat. Commun.* **10**, (2019).

15. Xiang, H. J. Origin of polar distortion in LiNbO3 -type 'ferroelectric' metals: Role of A -site instability and short-range interactions. *Phys. Rev. B - Condens. Matter Mater. Phys.* **90**, 1–7 (2014).

16. Jin, F. *et al.* Raman phonons in the ferroelectric-like metal LiOsO3. *Phys. Rev. B* **93**, 1–5 (2016).

17. Lo Vecchio, I. *et al.* Electronic correlations in the ferroelectric metallic state of LiOsO3. *Phys. Rev. B* **93**, 1–5 (2016).

18. Sim, H. & Kim, B. G. First-principles study of octahedral tilting and ferroelectric-like transition in metallic LiOsO3. *Phys. Rev. B - Condens. Matter Mater. Phys.* **89**, 1–5 (2014).

19. Kim, T. H. *et al.* Polar metals by geometric design. *Nature* **533**, 68–72 (2016).

20. Padmanabhan, H. *et al.* Linear and nonlinear optical probe of the ferroelectric-like phase transition in a polar metal, LiOsO3. *Appl. Phys. Lett.* **113**, (2018).

21. Yamamoto, S. *et al.* Element Selectivity in Second-Harmonic Generation of GaFeO3 by a Soft-X-Ray Free-Electron Laser. *Phys. Rev. Lett.* **120**, 1–9 (2018).

22. Lam, R. K. *et al.* Soft x-ray second harmonic generation as an interfacial probe. *Phys. Rev. Lett.* **120**, 23901 (2018).

23. Helk, T. *et al.* Table-top Nonlinear Extreme Ultraviolet Spectroscopy. *arXiv:2009.05151*, 1–16 (2020).

24. Schwartz, C. P. *et al.* Ångström-resolved interfacial structure in organic-inorganic junctions.

*arXiv:2005.01905*, 1–19 (2020).

25. Shwartz, S. *et al.* X-ray second harmonic generation. *Phys. Rev. Lett.* **112**, 1–5 (2014).

26. Owada, S. *et al.* A soft X-ray free-electron laser beamline at SACLA: The light source, photon beamline and experimental station. *J. Synchrotron Radiat.* **25**, 282–288 (2018).

27. Gulans, A. *et al.* Exciting: A full-potential all-electron package implementing density-functional theory and many-body perturbation theory. *J. Phys. Condens. Matter* **26**, (2014).

28. Perdew, J. P. & Zunger, A. Self-interaction correction to density-functional approximations for many-electron systems. *Phys. Rev. B* **23**, 5048–5079 (1981).

29. Sharma, S. & Ambrosch-Draxl, C. Second-harmonic optical response from first principles. *Phys. Scr.* **T109**, 128–134 (2004).

30. Kresse, G. & Furthmüller, J. Efficient iterative schemes for ab initio total-energy calculations using a plane-wave basis set. *Phys. Rev. B - Condens. Matter Mater. Phys.* **54**, 11169–11186 (1996).

31. Pascal, T. A. *et al.* Finite temperature effects on the X-ray absorption spectra of lithium compounds: First-principles interpretation of X-ray Raman measurements. *J. Chem. Phys.* **140**, (2014).

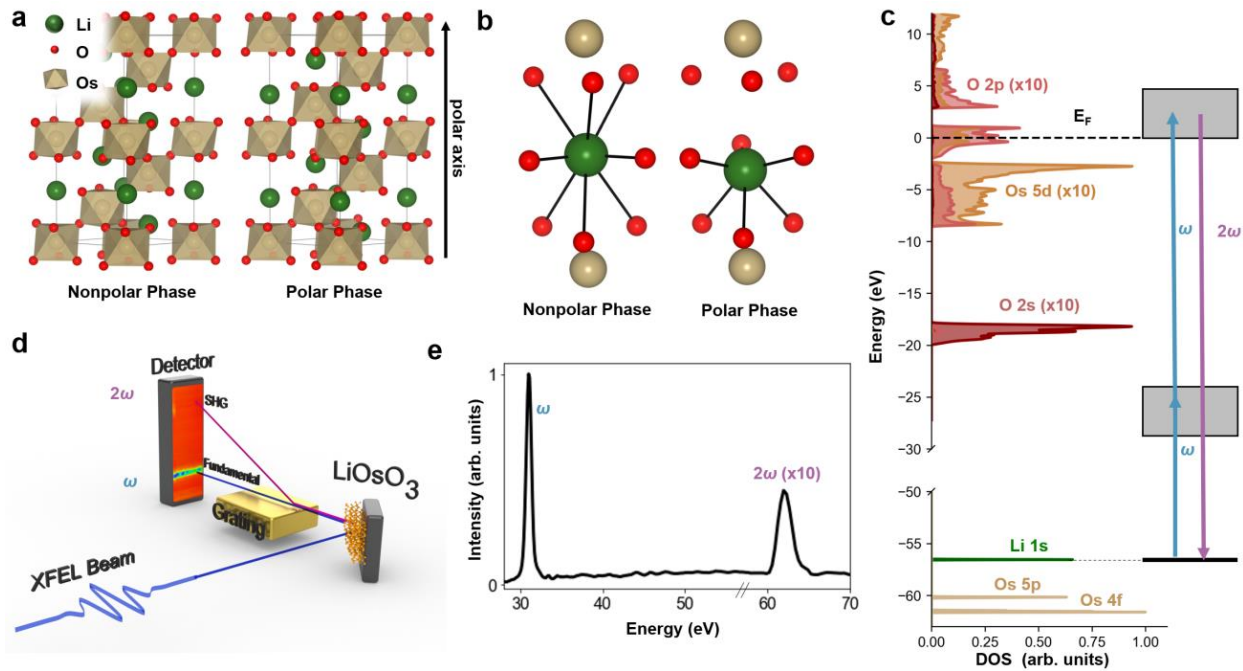

**Figure 1: The structures of LiOsO$_3$ in its nonpolar and polar phases, partial density of states calculations, and experimental scheme. a** The hexagonal unit cell of LiOsO$_3$ in its high-temperature nonpolar phase and low temperature polar phase with the polar c axis identified. **b** In the nonpolar phase, the Li atoms are symmetrically located between two OsO$_6$ octahedra from above and below and in plane with 3 nearby O atoms from adjacent OsO$_6$ octahedra. In the polar phase, an asymmetric displacement of Li atoms along the emergent polar c axis results in octahedrally coordinated Li atoms. **c** Partial density of states calculations. Near the Fermi level are predominantly states of Os 2p and Os 5d character. The XUV-SHG experiment involves core-level electron excitations at 28-33 eV photon energy ($\omega$) that are half-resonant with the Li 1s core state, resulting in second harmonic emission ($2\omega$). **d** Schematic of the experimental setup. The fundamental XUV beam is incident on the [120] plane of a LiOsO$_3$ crystal under 45° angle of incidence with respect to the surface normal. The reflected beam is spectrally dispersed by a grating onto a detector, enabling simultaneous measurement of signals at the fundamental and SHG frequencies. **e** A characteristic spectrum measured at an incident XFEL energy of 31 eV featuring peaks at both the fundamental and second harmonic.

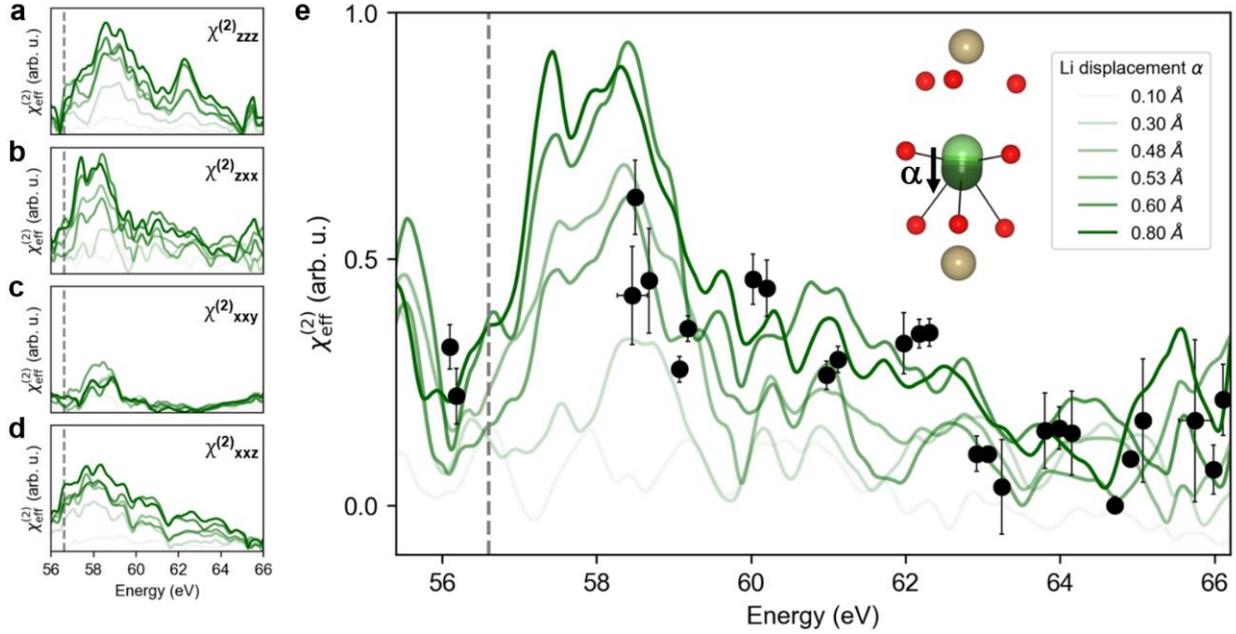

**Figure 2: The energy dependent nonlinear susceptibility across the Li K-edge. a-d** $\chi^{(2)}_{zzz}$, $\chi^{(2)}_{zxx}$, $\chi^{(2)}_{xxy}$ and $\chi^{(2)}_{xxz}$ across the Li K-edge. **e** The experimentally determined nonlinear susceptibility shown in as black points overlaid onto the calculated $\chi^{(2)}_{eff}(2\omega)$. For **a-e**, the dashed gray line at 56.6 eV corresponds to the onset of the Li K edge. darker green corresponds to larger Li displacement along the polar axis (*α*) of the LiOsO$_3$ polar phase as indicated by the inset in **e** and in the corresponding legend.

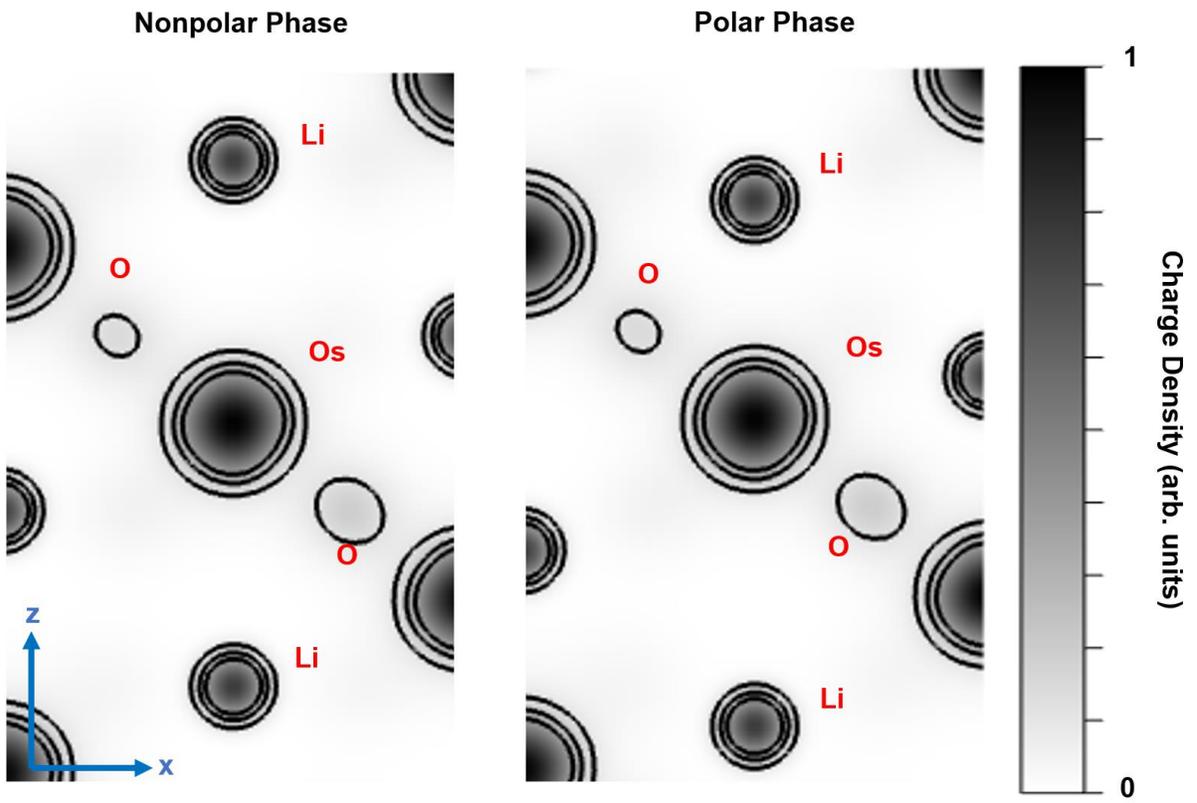

**Figure 3**: **Charge densities within the LiOsO$_3$ unit cell.** Charge densities along the polar axis reveal inversion symmetry is broken in the polar phase as Li atoms are displaced vertically along the polar z-axis. An applied electric field along the z-axis further increases the magnitude of this asymmetry, whereas an applied electric field in the xy-plane increases it non-trivially. The black contours are drawn at 0.15 intervals of charge density units from 0-0.6 arb. units.

# Supplementary Information: Direct observation of symmetry-breaking in a 'ferroelectric' polar metal

**Section S1: Extracting the nonlinear susceptibility from experimental data**
For each XFEL shot at a particular incident energy, the detector recorded three dominant features: the specular beam reflected off the grating, and the first diffraction order of the grating featuring a distinct signal for the fundamental XFEL photon energy and the second harmonic. In order to extract the shot-to-shot energy jitter of the XFEL, a Gaussian function was fit to each feature in each single shot image recorded. The peak positions of the fitted Gaussians were averaged and used as calibration points in the grating equation to convert from pixel on the detector to photon energy. For a given energy, each shot on the detector had a slight variation of about 4 pixels around the mean of a Gaussian fit to the spectral peak. Given the high number of shots (~30000) gathered (**Fig. S1.1**) at each photon energy each of the energy sub-bins had a sufficient number of shots to generate a plot of incident intensity vs. the second harmonic intensity (**Fig. S1.2**). To extract the $\chi^{(2)}(2\omega)$ value for each energy sub-bin, the shots classified in the sub-bin were put into 64 bins with respect to the incident fundamental intensity and subsequently averaged. This procedure was repeated for the second harmonic signal. The respective intensities for each bin were calculated by integrating the Gaussians using the trapezoidal rule. $\chi^{(2)}(2\omega)$ was extracted by linearizing the dependence shown in Eq.(1) of the main text by plotting the square of the fundamental intensity vs. the second harmonic intensity. A linear function was fit to the linear region of this plot, with the corresponding slope reporting on the $\chi^{(2)}(2\omega)$.

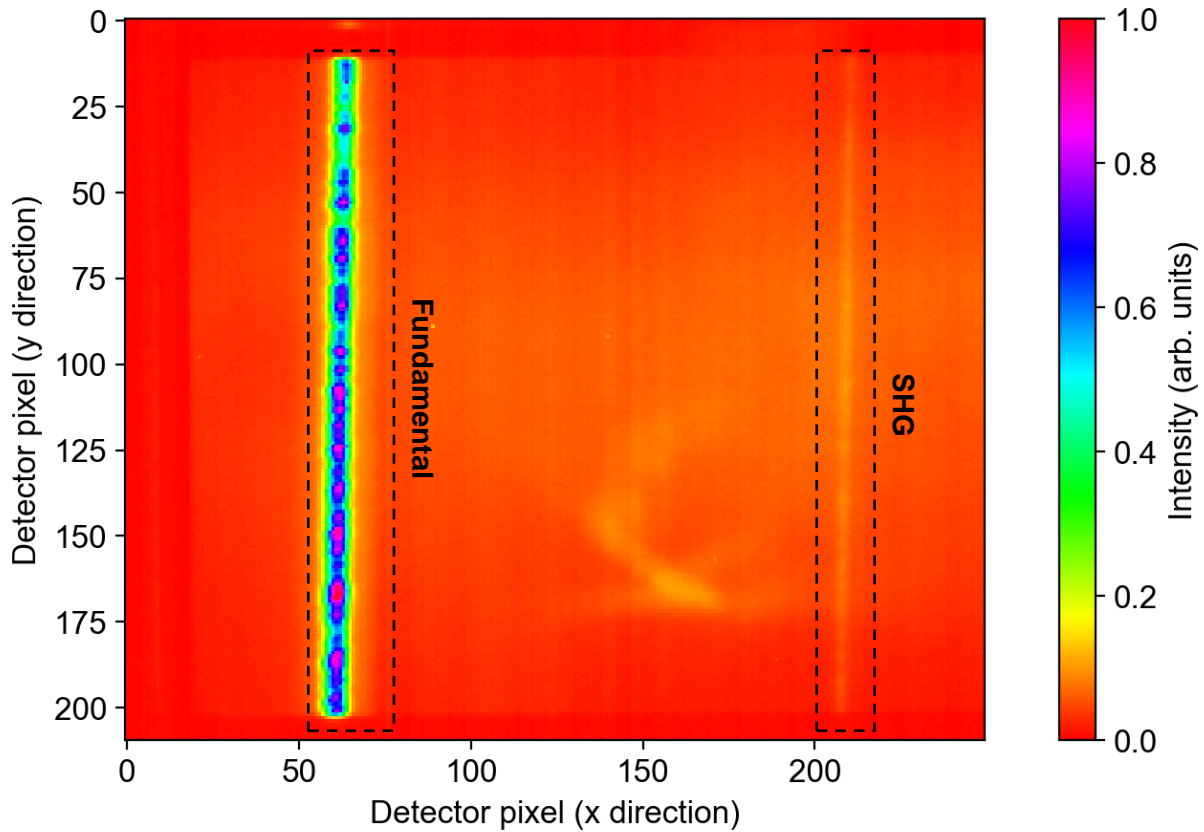

**Fig. S1.1. Representative image of spectrum captured with MCP detector.** All 30,000 shots at 31.5 eV were averaged together. In this image, no other modifications, including background corrections have been performed. Immediately apparent are the spectral features at the fundamental and second harmonic.

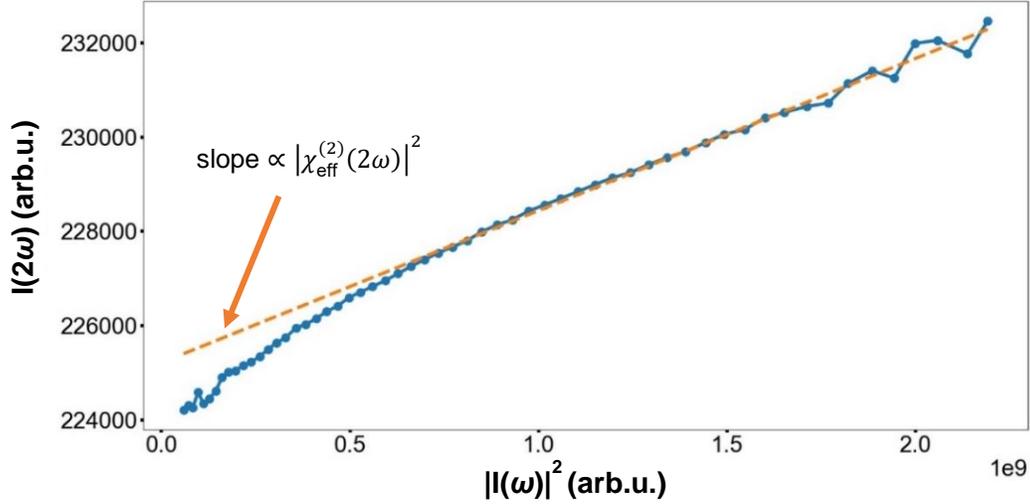

**Fig. S1.2. Linear fit *to |I(ω)|² vs. I(2ω)* to extract the nonlinear susceptibility $\chi_{\text{eff}}^{(2)}(2\omega)$**. After sorting all the spectra according to the pixel position of the fundamental on the detector x-axis, summing all recorded intensities in the detector y-direction, and performing a background correction, integrals under the fundamental and second harmonic spectral features were calculated. With these integrals corresponding to intensity (arb. units.) the effective nonlinear susceptibility could be extracted using Eq. (1) in the main text. A linearized version of Eq. (1) is plotted here, indicating the slope of the linear line fit to the most linear region of the data is proportional to $\left|\chi_{\text{eff}}^{(2)}(2\omega)\right|^2$.

### Section S2: The linear response

The absolute reflectance of $LiOsO_3$ was characterized at beamline 6.3.2 at the Advanced Light Source (ALS) at Lawrence Berkeley National Laboratory[1]. The reflectance from 20-70 eV was measured at four different incident angles with respect to the sample normal (10°, 20°, 30°, 45°). The imaginary part of the $LiOsO_3$ dielectric function, *κ(ω),* shown as a gray dashed line in **Fig. S2.1**, was extracted using the numerical algorithm as that used in the work by Kaplan *et al.*[2] The red and blue curves in the figure correspond to the calculated linear response of the centrosymmetric and non-centrosymmetric $LiOsO_3$ phases, respectively, that are shifted by +3.15 eV to match experiment. From the theory-calculated *κ(ω)*, we see the two phases have an almost identical response, which is in-line with previous calculations that show density of states (DOS) near the Fermi-level changes minimally between the two phases[3].

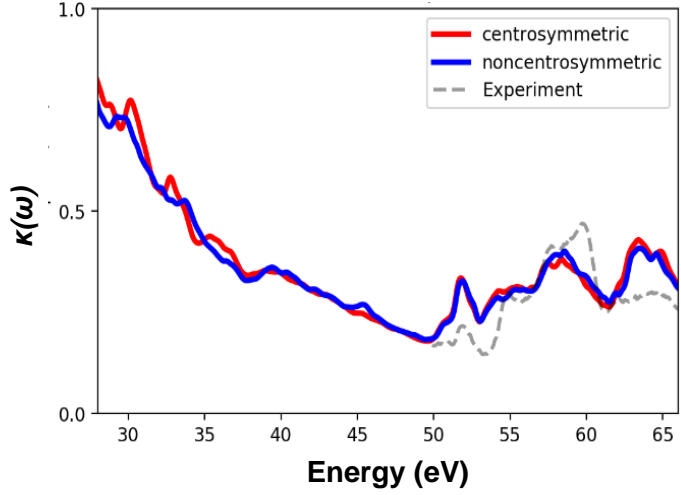

**Fig. S2.1. The imaginary part of the LiOsO3 dielectric function, *κ(ω)*.** Shown in red and blue are the calculated values of *κ(ω)* for the centrosymmetric and non-centrosymmetric phases. Shown with a gray dashed line is the experimentally extracted *κ(ω)* function, extracted from reflectivity data taken at the ALS. To align theory and experiment, a +3.15 eV shift in the theory x-axis was deemed necessary to best fit the data, as shown here.

### Section S3: Theoretical calculations of χ$^{(2)}$

The second-order nonlinear susceptibility was calculated according to the following equation

$$\chi^{(2)}_{ijk}(2\omega;\omega+\omega) = \frac{N}{\epsilon_0 \hbar^2} \sum_{nn'} \frac{\langle i \rangle_{gn} \langle j \rangle_{nn'} \langle k \rangle_{n'g}}{(\omega_{ng}-2\omega)(\omega_{n'g}-\omega)} + \frac{\langle j \rangle_{gn} \langle i \rangle_{nn'} \langle k \rangle_{n'g}}{(\omega_{ng}+\omega)(\omega_{n'g}-\omega)} + \frac{\langle j \rangle_{gn} \langle k \rangle_{nn'} \langle i \rangle_{n'g}}{(\omega_{ng}+\omega)(\omega_{n'g}+2\omega)}$$

where $\langle i \rangle, \langle j \rangle, \langle k \rangle$ represent the average electric dipole moments along the i,j, and k$^{th}$ Cartesian axes connecting the g, n', and n states which are indicated in the subscripts and denote the ground state, first excited state, and second excited state, respectively[4]. As can be seen from the first two terms on the right hand side of the equation, there is a strong resonant enhancement of the SHG signal when the frequency of incident light, ω, corresponds to the transition frequency between the ground and first excited states, ω$_{n'g}$. Within the *exciting* package, the simulation calculates the nonlinear response when a photon at the fundamental frequency is incident on the sample. In this experiment, the relevant part of the χ$^{(2)}$(2ω) spectrum thus includes energies at half of the measured SHG signal. Empty states are included in the ground state calculation to account for possible transitions to higher energy states, which results in Kohm-Sham eigenvalues related to the energy levels of the sample system.

### Section S4: Determining the χ$^{(2)}$$_{eff}$ response[4]

In contracted notation, $d_{ijk} = \frac{1}{2}\chi^{(2)}_{ijk}$, where the contracted matrix notation $d_{ijk} = d_{il}$ is given by the Voigt notation. The nonlinear susceptibility tensor for the R3c symmetry group has only 4 independent terms:

$$d_{il} = \begin{bmatrix} 0 & 0 & 0 & 0 & d_{15} & d_{16} \\ d_{16} & -d_{16} & 0 & d_{15} & 0 & 0 \\ d_{31} & d_{31} & d_{33} & 0 & 0 & 0 \end{bmatrix} \quad (1)$$



This matrix defines the polar axis along the z axis, while the experiment was performed with the polar axis perpendicular to the z-axis. To account for this, the $d_{il}$ tensor was rotated by 90° about the y-axis to give the rotated $d'_{il}$ tensor:

$$d_{il} = \begin{bmatrix} -d_{33} & -d_{31} & -d_{31} & 0 & 0 & 0 \\ 0 & -d_{16} & -d_{16} & 0 & 0 & d_{15} \\ 0 & 0 & 0 & -d_{16} & -d_{15} & 0 \end{bmatrix} \quad (2)$$

With this tensor, the second-order polarization response of the medium was calculated:

$$P_i(2\omega) = 2\varepsilon_0 d'_{il} E_j(\omega) E_k(\omega) \quad (3)$$

where i,j,k correspond to the x,y,z coordinate axes. The geometry of the experiment as shown in **Figure S4.1** is such that the incident beam makes an angle θ=45° with the z-axis and an angle φ=240° with the x-axis when projected into the xy-plane. The electric field of the incident photon at the fundamental frequency can then be decomposed into

$$\begin{bmatrix} E_x(\omega) \\ E_y(\omega) \\ E_z(\omega) \end{bmatrix} = \begin{bmatrix} -E_0 \sin\theta \cos\phi \\ -E_0 \sin\theta \sin\phi \\ -E_0 \cos\theta \end{bmatrix} = \frac{E_0}{\sqrt{2}} \begin{bmatrix} 1/2 \\ \sqrt{3}/2 \\ -1 \end{bmatrix} \quad (4)$$

where $E_0$ is the amplitude of the incoming electric field. Substituting equations (2) and (4) into (3) gives the induced polarization

$$\vec{P} = \begin{bmatrix} P_x(2\omega) \\ P_y(2\omega) \\ P_z(2\omega) \end{bmatrix} = -\epsilon_0 \frac{E_0^2}{4} \begin{bmatrix} d_{33} + 7 d_{31} \\ -2\sqrt{3} d_{15} - d_{16} \\ -4\sqrt{3} d_{16} - 4 d_{15} \end{bmatrix} \quad (5)$$

To determine the polarization the detector sees, the polarization vector needs to be decomposed into two parts that carry polarization parallel and perpendicular to the wave vector of the reflected beam. Since a traveling photon can only have transverse polarization, the detector only measures the $\| P_\perp \|$ component. In spherical coordinates, the outgoing reflected beam lies along the vector $(r, \theta, \phi) = (1, 45°, 60°)$. In Cartesian coordinates, this is the vector $\vec{R} = \left( \frac{1}{2\sqrt{2}}, \frac{\sqrt{3}}{2\sqrt{2}}, \frac{1}{\sqrt{2}} \right)$ with a norm of 1. The projection of $\vec{P}$ onto $\vec{R}$ is

$$P_\| = \text{proj}_{\vec{R}} \vec{P} = (\vec{R} \cdot \vec{P}) \vec{R} = \left( \frac{P_x}{8}, \frac{3 P_y}{8}, \frac{4 P_z}{8} \right) \quad (6)$$

The perpendicular component is then

$$\begin{aligned} P_\perp &= \vec{P} - \vec{P}_\| = \left( \frac{7 P_x}{8}, \frac{5 P_y}{8}, \frac{4 P_z}{8} \right) \\ &= -\frac{\epsilon_0 E_0^2}{32} \left( 7 d_{33} + 49 d_{31}, -10\sqrt{3} d_{15} - 5 d_{16}, -16\sqrt{3} d_{16} - 16 d_{15} \right) \\ \| P_\perp \| &= \frac{\epsilon_0 E_0^2}{32} (49 d_{33}^2 + 2401 d_{31}^2 + 556 d_{15}^2 + 793 d_{16}^2 + 686 d_{31} d_{33} + 612\sqrt{3} d_{15} d_{16})^{1/2} \end{aligned} \quad (7)$$



therefore

$$d_{\text{eff}} = \frac{1}{32}(49d_{33}^2 + 2401d_{31}^2 + 556d_{15}^2 + 793d_{16}^2 + 686d_{31}d_{33} + 612\sqrt{3}d_{15}d_{16})^{1/2} \tag{8}$$

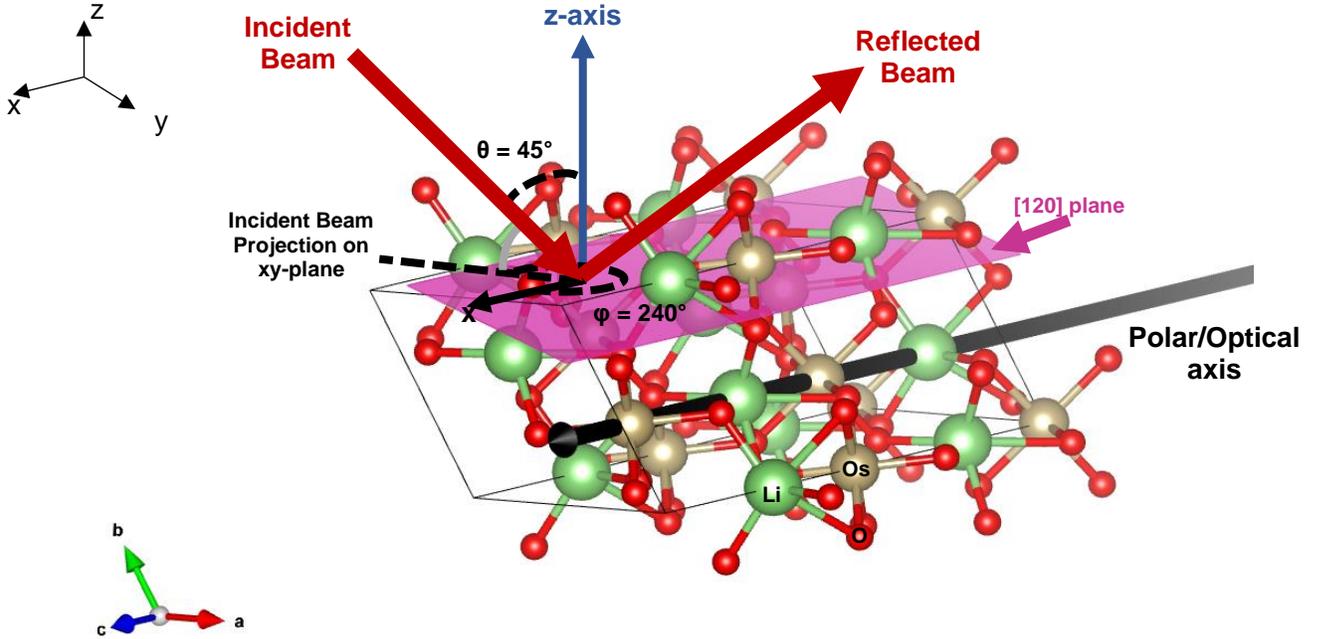

**Fig. S4.1**. **Experimental geometry of XUV-SHG experiment.** In the experiment, the XFEL beam was incident at 45° with respect to the [120] plane of LiOsO$_3$ (shown in pink). The projection of the incoming beam onto the xy-plane, as defined in the figure, made a 240° angle with the LiOsO$_3$ polar axis. The detector was placed at 45° with respect to the sample normal.

**Section S5: Neglecting the transient response of the FEL during the duration of the pulse**
It is important to consider whether the Li ion could potentially be displaced within the envelope of the incident FEL pulse by the FEL pulse itself. If this were the case, the SHG signal would probe a transient excited state, rather than the ground state. Such a process would occur on a few-femtosecond timescale. However, simulations indicate that within the 30 fs of the XFEL pulse, the Li-ion would move by approximately 0.02 Å at $T$ = 62 K. Given the length scales of interest for the experiment were an order of magnitude larger, the transient response of the sample to the XFEL pulse could be reasonably neglected (**Fig. S5.1**).

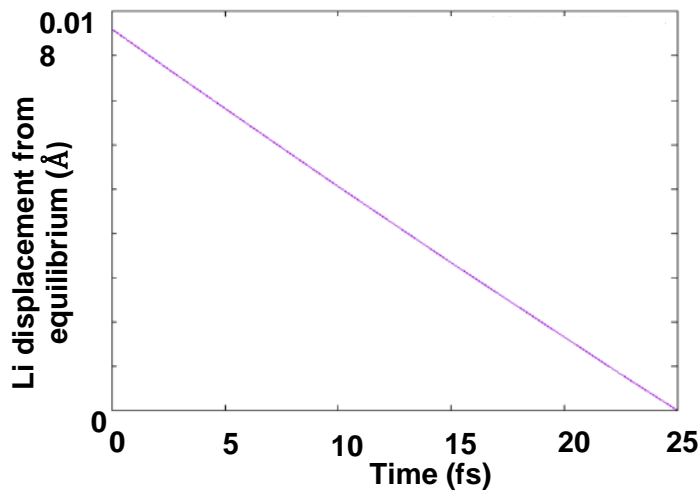

**Fig. S5.1. The transient response of the Li-ion position with respect to the XFEL pulse**. On the timescale of the 30 fs XFEL pulse, the Li-ion is expected to displace by about 0.02 Å. This length scale is an order of magnitude smaller than that of interest in the experiment. As such, the transient response of the XFEL could be reasonably neglected.

**Section S6: Partial Density of States (PDOS) Calculations**

The density of states calculation of LiOsO$_3$ was performed in exciting. The effective k-point mesh size for Brillouin zone integration *ngrdos* was set to 1000. The energy range was chosen from -1.9 to 2.0 Hartree. The spherical harmonic basis set was transformed into site symmetries which gave the physical contribution of the DOS in term of angular quantum number or the spdf orbital characteristic. The PDOS near the Li K-edge includes contributions from Li 1s orbitals, Os 5p orbitals, and Os 4f orbitals at approximately –56.6, -60.2, and -61.6 eV, respectively. The PDOS at the Fermi level is dominated by contributions from Os 5d and O 2p orbitals and significantly smaller contributions from O 2s, Li 2s, Li 2p, Os 5s, and Os 5p orbitals (**Fig. S6.1**). The selection rules for SHG, however, only allow transitions between states for which $\Delta l = 0, \pm 2$. Given the resonant enhancement expected for incident photons at the Li K-edge, transitions originating from Os 4p and 4f orbitals are not expected to complicate the signal since transitions from s-type orbitals to p- and f-type orbitals are forbidden and as shown in Fig. 3 of the main text, Os atoms are in a predominantly centrosymmetric environment, for which SHG is not allowed. Of the transitions from Li 1s states that are allowed, O 2s orbitals are below the Fermi level and are not only occupied but also inaccessible with incident photons within the 28-33 eV range used. Furthermore, contributions of Li 2s orbitals to the TDOS are 2 orders of magnitude smaller than those of Os 5d orbitals. It can therefore be reasonably assumed that the SHG signal is selective to the Li coordination environment and represents transitions mainly from Li 1s to Os 5d orbitals.

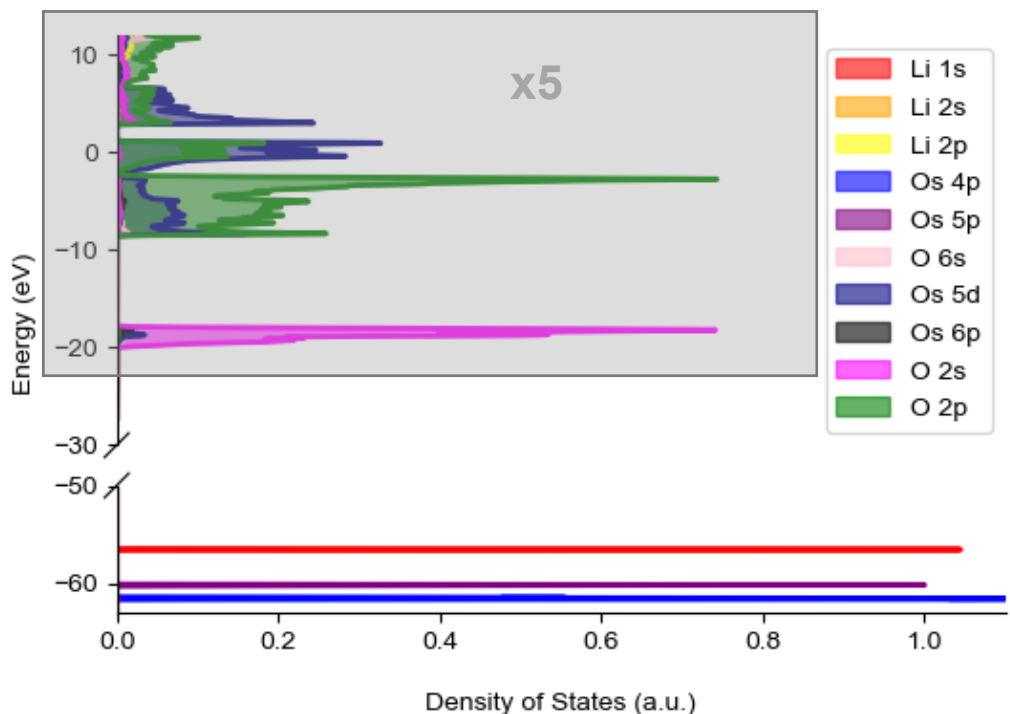

Fig. S6.1. Partial density of states of LiOsO$_3$ determined by DFT.

**Section S7: Theoretical Calculation of $\chi^{(2)}_{eff}$ without Li 1s response**

To verify that origin of second harmonic response of LiOsO$_3$ is due to the displacement of Li atoms, we explicitly excluded the Li 1s electron from the SCF calculation. In this way the nonlinear response calculation would not include the signal from Li 1s signal. The effective second order nonlinear susceptibility is shown in **Fig. S7.1**. We observed that the majority of the nonlinear response is contributed by the Li 1s electrons. The signal from other electrons while present is heavily overshadowed by that of Li 1s electrons. Therefore, the signal we're observing primarily originate from the Li 1s which in turn is enhanced by the displacement of Li atoms from centrosymmetric positions.

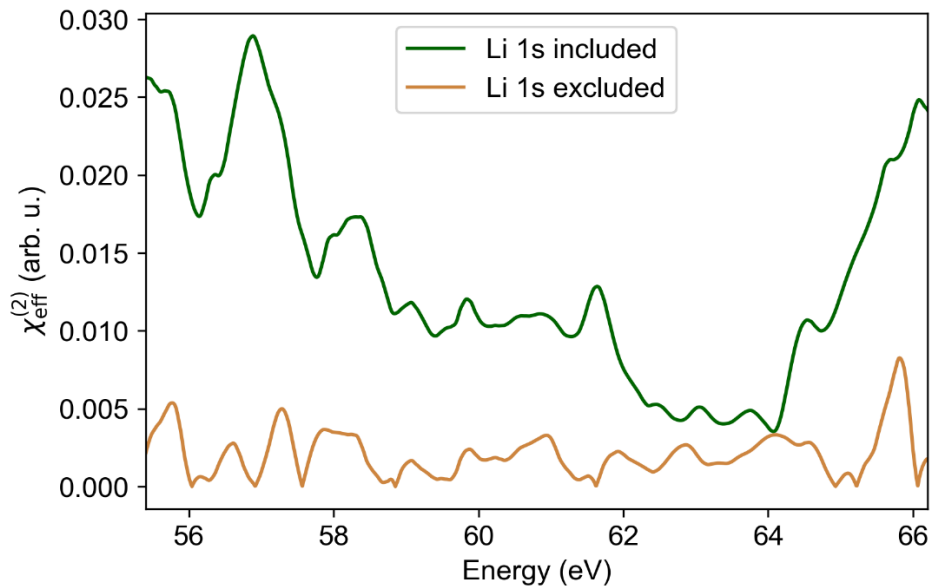

**Fig. S7.1: $\chi^{(2)}_{eff}$ across the Li K-edge comparison between systems with and without Li 1s electron contribution.** It can be seen that the response at 57 eV is due to the Li 1s core, as this response is not present when the Li 1s core is absent.

**Section S8: Effect of OsO$_6$ octahedral rotation to $\chi^{(2)}_{eff}$**

From the neutron diffraction experiment of LiOsO$_6$, it was reported that polar and nonpolar phase OsO$_6$ octahedral rotated to allow better hybridization due to the difference between Li position in the two phases[5]. We calculate the effect on the $\chi^{(2)}_{eff}$ due to different OsO$_6$ orientation. From the result shown in Fig. S8, the rotation of OsO$_6$ has minimal effect on the second harmonic response of the LiOsO$_3$, all the features between the two phases are very similar. Due to the sensitivity of SHG, the change in the environment of the osmium having minimal effect to $\chi^{(2)}_{eff}$ suggests that the main signal contribution arises from the Li atoms.

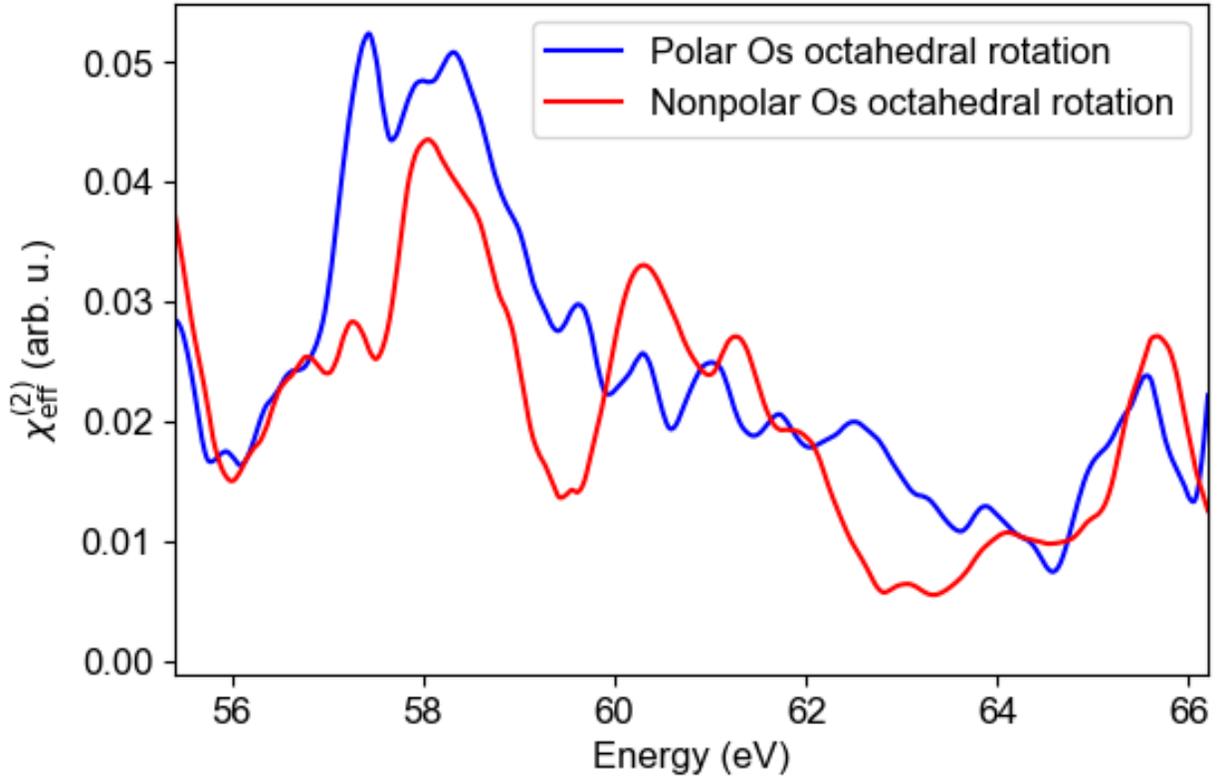

**Fig. S8.1: $\chi^{(2)}_{eff}$ across the Li K-edge comparison between systems polar and nonpolar OsO$_6$ geometrical orientation.**

**References**
1. Underwood, J. H. *et al.* Calibration and standards beamline 6.3.2 at the Advanced Light Source. *Rev. Sci. Instrum.* **67**, 3372–3372 (1996).
2. Kaplan, C. J. *et al.* Retrieval of the complex-valued refractive index of germanium near the M 4,5 absorption edge . *J. Opt. Soc. Am. B* **36**, 1716 (2019).
3. Sim, H. & Kim, B. G. First-principles study of octahedral tilting and ferroelectric-like transition in metallic LiOsO3. *Phys. Rev. B - Condens. Matter Mater. Phys.* **89**, 1–5 (2014).
4. Boyd, R. W. *Nonlinear Optics (3rd Edition)*. (Academic Press, 2008).
5. Shi, Y. *et al.* A ferroelectric-like structural transition in a metal. *Nat. Mater.* **12**, 1024–1027 (2013).